\documentclass{article}
\usepackage{times,fancyhdr}
\usepackage{graphics}
\usepackage{fancyhdr}
\usepackage{amsmath}
\usepackage{amsfonts}
\usepackage{amssymb}
\usepackage{cite}
\usepackage{graphicx}
\newtheorem{theorem}{Theorem}

\newtheorem{example}[theorem]{Example}

\newtheorem{remark}[theorem]{Remark}

\begin{document}

\title{The motion of stars in galaxies and the gravitational time dilatation}
\author{Emmanuel Moulay \thanks{E-mail: emmanuel.moulay@univ-poitiers.fr}\\
\\ Xlim (UMR-CNRS 6172) - D\'{e}partement SIC\\
Universit\'{e} de Poitiers - France}
\date{}

\maketitle

\begin{abstract}
This article deals with the problem of the motion of stars in galaxies. By using the Newton's theory combined with a gravitational time dilatation for the weak gravitational field, it is possible to give a solution without using the dark matter.
\end{abstract}

\section{Introduction}
Since the observations of Zwicky in \cite{Zwicky33}, we wonder why the motion of stars in galaxies do not corroborate the gravitational effects. The main explanation is the \emph{dark matter} which is hypothetical matter of the standard cosmological model $\Lambda$-CDM that is undetectable by its emitted radiation, but whose presence can be inferred from gravitational effects on visible matter \cite{Papantonopoulos07}. However, such a theory raises some problems \cite{BlanchetBis09}. There exists an alternative theory to the dark matter which is the \emph{Modified Newtonian dynamics} MOND of Milgrom who has developed a change in the Newton's law  \cite{Milgrom83,Milgrom83bis,Milgrom83Ter}. A comparison between these two approaches is developed in \cite{Blanchet09} by Blanchet who has also proposed a new kind of dark matter: the dipolar dark matter \cite{BlanchetBis09}.\\
The main goal of this paper is to provide another solution for the problem of the motion of stars in galaxies by using a new gravitational time dilation for the weak gravitational field combined with the Newton's theory.\\
The paper is organized as follows. The problem of the motion of stars in galaxies is recalled in Section \ref{SecDM} and the solution involving a new gravitational time dilation for the weak gravitational field is addressed in Section \ref{SecTD}. Finally, a conclusion is given in Section \ref{SecConc}

\section{The problem of the motion of stars in galaxies}\label{SecDM}

The galaxies are characterized by a very weak gravitational field. So, such astrophysical systems must be correctly described by the Newton's theory of gravitation.

At a distance $d$ of the center of a galaxy, the \emph{Newton's gravitational force} $F_N$ a star undergoes is given by:
\begin{equation}
F_N(d)=G\;\frac{M_g \; M_s}{d^2}
\end{equation}
where $G$ is the gravitational constant, $M_g$ the mass of the galaxy and $M_s$ the mass of the star. Using the Newton's law of dynamics gives the following acceleration for the star:
\begin{equation}
a(d)=G\;\frac{M_g}{d^2} \; m.s^{-2}.
\end{equation}
Since the equation that relates the velocity to the acceleration for a circular orbit is given by:
\begin{equation}
a=\frac{v^2}{d},
\end{equation}
we have the following velocity for a star located at the distance $d$ from the center of a galaxy:
\begin{equation}\label{Vit}
v(d)=\sqrt{\frac{G\;M_g}{d}} \; m.s^{-1}.
\end{equation}
It seems to depend on $d$.

\begin{figure}[h]
  \begin{center}
  \includegraphics[width=6cm,height=5cm]{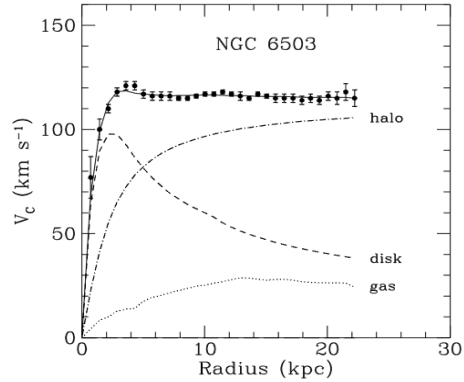}\\
  \caption{Galaxy rotation curve}\label{vitesse}
  \end{center}
\end{figure}

Instead of decreasing asymptotically to zero as the effect of gravity wanes, the observed distribution curve remains flat, showing the same velocity at increasing distances from the bulge.

\section{The solution of the gravitational time dilation}\label{SecTD}

At this stage, either we suppose that the previous calculus is correct or not. In this paper, we postulate that the Newton's dynamics are correct. We also know that the restrictions on variations in the gravitational constant are strict \cite{Uzan03}. Nevertheless, we want to provide a solution for the problem of the motion of stars in galaxies without using the assumption of the dark matter.

The Newton's theory is only an approximation of relativity. Something happens in relativity which is not taken into account: this is the gravitational time dilation. We mean the gravitational time dilation due to the changing gravitational field, not the relativistic time dilatation due to the velocity of the stars which is not significant. Let us recall that the \emph{relativistic time dilation} induced by the velocity is given by:
\begin{equation}
\Delta T = \sqrt{1-\frac{v^2}{c^2}} \; \Delta T'
\end{equation}
where $c$ is the speed of light and $v$ the relative velocity between the observer whose flow of time is given by $\Delta T'$ and the moving clock whose flow of time is given by $\Delta T$ \cite[Section 2.4]{Gron07}.

Due to the huge value of $d$, we know that the gravitational field is very weak. The problem of the motion of stars in galaxies could indicate that the gravitational time dilatation cannot be neglected for the weak gravitational field. However, it is rather difficult to evaluate this gravitational time dilation due to the changing weak gravitational field which leads to the motion of stars in galaxies. Indeed, gravitational time dilation effects for the solar system and the Earth have been evaluated from the starting point of an approximation to the \emph{Schwarzschild solution} to the Einstein's field equation. It is given for all $d,d'>R_S$ by:
\begin{equation}
\Delta T(d)=\sqrt{\frac{1-\frac{R_s}{d}}{1-\frac{R_s}{d'}}} \; \Delta T(d')
\end{equation}
where $R_S= \frac{2 G M}{c^2}$ is the Schwarzschild radius \cite[Page 219]{Gron07}. If we use the Schwarzschild solution to describe the motion of stars in galaxies, such a gravitational time dilatation is not significant. However, if the Schwarzschild solution is an exact solution of the Einstein's field equation which correctly describes the strong gravitational field outside a spherical non-rotating mass, we cannot be sure that the same is true for the weak gravitational field of galaxies. We have no relevant way for evaluating this gravitational time dilation in galaxies, due in particular to the special distribution of matter in galaxies and above all, due to the fact that the Einstein's gravitational field has never been tested for the weak gravitational field.

Let $d\gg0, \; d'\gg0$ such that $a(d)\ll a_0, \; a(d')\ll a_0$ with $a_0 \sim 1.2 \times10^{-10} m.s^{-2}$ the constant given in the MOND theory. Suppose that the gravitational time dilation, induced by the weak gravitational field for stars located at $d, \; d'$ in galaxies, is as follows:
\begin{equation}\label{time}
\Delta T(d) = \sqrt{\frac{d}{d'}}\; \Delta T(d').
\end{equation}
We draw the reader's attention to the fact that the value of $\sqrt{\frac{d}{d'}}$ is rather weak because this is the ratio of the huge distances from the center of the galaxy which is taken into account. Equation (\ref{time}) is consistent with the fact that the time flow slows down when the gravitational field decreases, i.e. when the distance to the center of the galaxy increases.  Indeed, one second at the distance $d$ of the center of the galaxy corresponds $\sqrt{\frac{d'}{d}}$ seconds at the distance $d'$ of the center of the galaxy. For an observer located at $d$, the observed velocity at $d'$ satisfies:
\begin{equation}
v_{obs}(d')=v(d')\sqrt{\frac{d'}{d}}=v(d)
\end{equation}
which is in accordance with the observations given by Figure \ref{vitesse}. Indeed, the observer, located at the distance $d\gg0$, observers the same velocity for the stars located at all distances $d'\gg0$.

\begin{example}
Let us consider the motion of the Sun and the one of another star into the Milky Way. We suppose that the star is 10 times closer to the galactic center than the Sun. Recall that the distance from the Sun to the galactic center is now estimated at $d'_{su}=26,000 \; ly$, so the distance from the star to the galactic center is $d_{st}=2,600 \; ly$. The gravitational time dilatation (\ref{time}) is given by:
\begin{equation}
\Delta T(d_{st})=\sqrt{\frac{1}{10}} \; \Delta T(d'_{su}) \approx 0.316 \; \Delta T(d'_{su}).
\end{equation}
This example shows that the problem of the motion of stars in our galaxy can be solved by using a gravitational time dilation of about 3 times with respect to the flow of time of the weak gravitational field where the Sun orbits the Milky Way.
\end{example}

\begin{remark}
We can infer that if the gravitational time dilatation of the weak gravitational field is responsible for the problem of the motions of stars in galaxies, the same could be true for the problem of the \emph{dark energy} which is the hypothetical form of energy that tends to increase the rate of expansion of the universe \cite{Papantonopoulos07}. In such a case, the weak gravitational field becomes weaker in the intergalactic space and the gravitational time dilatation between the galaxies increases. If the flow of time elapses quickly in the intergalactic space, it implies that the rate of expansion between the galaxies increases.
\end{remark}

This article raises the problem of finding a relativistic theory for the weak gravitational field which would be consistent with the gravitational time dilation (\ref{time}), but also with the other observations such as gravitational lenses, considering that the Einstein's field equation is the one of the strong gravitational field. We can mention the \emph{tensor-vector-scalar gravity (TeVeS)} developed by Bekenstein and Sanders in \cite{Bekenstein04,Sanders05} as an attempt to define such a theory. Finally, the \emph{Weyl tensor equation} is also an appropriate tool for developing a relativistic theory for the weak gravitational field and also a complete Riemannian theory of gravitation \cite[Section 4.1]{Hawking75}.

\section{Conclusion}\label{SecConc}
Even if the $\Lambda$-CDM model is the main theory, we cannot rule out other possibilities, as long as we have no direct evidence of the existence of dark matter. The gravitational time dilatation for the weak gravitational field is one of the possibility to explain the motion of stars in galaxies.

\bibliographystyle{plain}
\bibliography{motion}

\end{document}